\documentclass[super]{rsc-layout}
\journalname{Soft Matter}

\usepackage{graphicx,amsmath,amssymb,textcomp}
\usepackage[countmax]{subfloat}
\usepackage{txfonts,times}
\graphicspath{{figures/}}

\renewcommand\epsilon{\varepsilon}
\renewcommand\phi{\varphi}
\renewcommand\theta{\vartheta}
\renewcommand\vec[1]{\textrm{\bfseries #1}}

\newcommand\diff{\mathrm{d}}
\newcommand\dotprod{\boldsymbol{\cdot}}
\newcommand\expect[1]{\left\langle{#1}\right\rangle}

\newcommand\dw{{d_\text{w}}}
\newcommand\df{{d_\text{f}}}
\newcommand\ds{{d_\text{s}}}

\newcommand\eq[1]{Eq.~\eqref{eq:#1}}
\newcommand\fig[1]{Fig.~\ref{fig:#1}}

\makeatletter
\def\allowhyphens{\penalty\@M \hskip\z@skip}
\makeatother

\begin{document}

\title{Anomalous transport resolved in space and time by fluorescence \mbox{correlation} spectroscopy}

\author{%
Felix H\"of{}ling\footnotemark[1]\footnotemark[2]%
{\renewcommand\thefootnote{\fnsymbol{footnote}}%
\footnotemark[1]}%
,
Karl-Ulrich Bamberg\footnotemark[2]%
, and
Thomas Franosch\footnotemark[2]\footnotemark[3]%
}

\date{\today}

\abstract{%
A ubiquitous observation in crowded cell membranes is that molecular transport
does not follow Fickian diffusion but exhibits subdiffusion.
The microscopic origin of such a behaviour is not understood and highly debated.
Here we discuss the spatio-temporal dynamics for two models of subdiffusion:
fractional Brownian motion and hindered motion due to immobile obstacles.
We show that the different microscopic mechanisms can be distinguished using
fluorescence correlation spectroscopy  (FCS)
by systematic variation of the confocal detection area.
We provide a theoretical framework for space-resolved FCS by
generalising FCS theory beyond the common assumption of spatially Gaussian transport.
We derive a master formula for the FCS autocorrelation function,
from which it is evident that the beam waist of an FCS experiment is
a similarly important parameter as the wavenumber of scattering experiments.
These results lead to scaling properties of the FCS correlation
for both models, which are tested by \emph{in silico} experiments.
Further, our scaling prediction is compatible with the FCS half-value times
reported by Wawrezinieck \emph{et al.} [Biophys. J. 89, 4029 (2005)]
for \emph{in vivo} experiments on a transmembrane protein.}


\maketitle

\bgroup
\renewcommand\thefootnote{\alph{footnote}}
\footnotetext[1]{%
Rudolf Peierls Centre for Theoretical Physics, 1~Keble Road, Oxford OX1 3NP, England, United Kingdom}
\footnotetext[2]{%
Arnold Sommerfeld Center for Theoretical Physics (ASC) and Center for Nano Science (CeNS), Fakult{\"a}t f{\"u}r Physik, Ludwig-Maximilians-Universit{\"a}t M{\"u}nchen, Theresienstra{\ss}e 37, 80333 M{\"u}nchen, Germany}
\footnotetext[3]{%
Institut f\"ur Theoretische Physik, Friedrich-Alexander-Universit\"at Erlangen--N\"urnberg, Staudtstra{\ss}e 7, 91058 Erlangen, Germany}
\egroup

\footnotetext[1]{Present address: Max-Planck-Institut f\"ur Metallforschung,
Heisenbergstra{\ss}e 3, 70569
Stuttgart and Institut f\"ur Theore\-tische und Angewandte Physik, Universit\"at
Stuttgart, Pfaffenwaldring 57, 70569 Stuttgart, Germany}
\setcounter{footnote}{1}



\section*{Introduction}

Measurement of molecular transport at the subcellular level can provide important information on both physiological mechanisms and physical interactions that drive and constrain biochemical processes.
The obstructed motion of biomolecules in living cells displays anomalous transport including subdiffusion, which was established in the past decade by numerous experiments
applying techniques with labeled particles.
Nevertheless, the interpretation of the collected data remains often controversial
and the origin of the subdiffusive behaviour is highly debated~\cite{Szymanski:2009,Magdziarz:2009,He:2008,Lubelski:2008,Saxton:2007,Sung:2006}.
Crowded environments like cellular membranes contain structures on many length scales, and further progress depends on experimental techniques that resolve transport on these different scales.
Such spatio-temporal information is needed to test and refine models of anomalous transport.

One widespread technique for the investigation of molecular transport is fluorescence correlation spectroscopy (FCS), which follows the motion of fluorescently labeled molecules with high temporal resolution~\cite{Hess:2002,Schwille:2001}.
This mesoscopic, local method consists of collecting the fluorescent light from a steadily illuminated volume or area and autocorrelating its intensity fluctuations.
An important parameter of FCS measurements is the beam waist of the
illumination laser. While experimental setups in the past were constrained to a fixed value,
recent technological advancements allow large variations of the confocal detection area to gather spatial information~\cite{Gielen:2009, Wawrezinieck:2005, Masuda:2005, Wenger:2007, Hell:2007, Eggeling:2009}.
By z-scan FCS~\cite{Gielen:2009, Wawrezinieck:2005}
or beam expanders~\cite{Masuda:2005},
focal radii can be varied between 200\,nm and 500\,nm,
but measurements at the nanoscale became possible by introducing nano-apertures (75\,nm to 250\,nm)~\cite{Wenger:2007}.
Only lately, a far-field optical nanoscopic method named stimulated emission depletion (STED) fluorescence correlation microscopy was developed to beat the diffraction limit~\cite{Hell:2007},
allowing focal radii to span almost a decade down to 15\,nm (Ref.~\citenum{Eggeling:2009}).

The subdiffusive motion of macromolecules in crowded cells and membranes was studied extensively by FCS experiments~\cite{Guigas:2007a, Banks:2005, Weiss:2003, Avidin:2010}
and was complemented in real-space by single-particle tracking~\cite{Selhuber-Unkel:2009, Golding:2006, Kusumi:2005, Tolic-Norrelykke:2004}.
For Fickian diffusion, the mean-square displacement
grows linearly in time, $\delta r^2(t)=4 Dt$ in two dimensions with diffusion constant~$D$.
Then, the decay of the FCS autocorrelation function obeys
\begin{equation}
G(t) = \frac{1}{N} \frac{1}{1+ t/\tau_D} \, ,
\label{eq:fcs_normal}
\end{equation}
where $\tau_D = w^2/4 D$ denotes the dwell time and $N$ the average number of labeled molecules in the illuminated area.%
\footnote{%
We restrict the discussion to two dimensional systems relevant for membranes, where
focus distortions are negligible; in three dimensions, the asphericity of the
illumination volume renders the formulae more cumbersome.
We also ignore effects due to the photophysics of the dye molecules, which are relevant
at very short time scales only.
This does not effect the generality of our discussion nor any of our conclusions.}
These equations are no longer valid for anomalous transport. Introducing the walk dimension $\dw$, subdiffusion is characterised by $\delta r^2(t)\sim t^{2/\dw}$,
and FCS experiments are often rationalised by
\begin{equation}
G(t) = \frac{1}{N} \frac{1}{1+ (\Gamma t)^{\alpha}}
\label{eq:fcs_anomalous}
\end{equation}
upon fitting 
$N$, $\Gamma$, and $\alpha$.
It is usually and tacitly anticipated that both exponents coincide, $\alpha=2/\dw$.

Here, we provide a theoretical framework for space-resolved FCS.
Relating the FCS function $G(t)$ to the intermediate scattering function,
we generalise the conventionally used fit models and connect FCS to time-resolved scattering techniques.
If the beam waist is considered an adjustable experimental parameter similar to the scattering angle, FCS is turned into a valuable tool for the investigation of complex
and in particular anomalous transport.
The new approach greatly facilitates \emph{in silico} experiments: for two models of subdiffusion,
we show how spatio-temporal information on the tracer dynamics can be obtained and used to
distinguish different mechanisms as the origin of anomalous transport.

\section*{Theory}

\paragraph*{Generalised FCS theory.}
Let us briefly revisit the theory underlying the FCS technique~\cite{BernePecora:DynamicLightScattering,Schwille:2001}; we specialise to two dimensions for simplicity.
The detection area is illuminated by a laser beam with intensity profile $W(\vec r)$.
The fluorescent light depends on the fluctuating, local concentration $c(\vec r,t)$ of labeled molecules in the laser focus.
Thus, the intensity collected at the detector is a spatially weighted average,
$I(t) \propto \int\!\diff^2 r\, W(\vec{r})\,  c(\vec{r},t)$.
The output of the FCS experiment is the time-autocorrelation function of the intensity fluctuation $\delta I(t) = I(t) - \expect{I}$ around the mean intensity.
It is conventionally normalised as $G(t) =  \expect{\delta I(t)\,\delta I(0)}/ \expect{I}^2$;
proper normalisation would be achieved by multiplication with $N=\expect{I}^2/\expect{\delta I^2}$. Introducing spatial Fourier transforms, one arrives at the representation
\begin{equation}
 G(t;w) = \frac{1}{N} \frac{\int\! \diff^2 q\, |W(\vec{q})|^2\, S(\vec{q},t)}
 {\int \diff^2 q\, |W(\vec{q})|^2\, S(\vec{q},t=0)} \, ,
\label{eq:fcs_integrals}
\end{equation}
where $S(\vec{q},t) = \int\! \diff^2 r \,
\exp(\text{i} \vec{q} \dotprod \vec{r})
\expect{\delta c(\vec{r},t)\, \delta c(\vec{0},0)}$
is known as the intermediate scattering function
and $W(\vec{q})$ denotes the Fourier transform of the intensity profile $W(\vec{r})$.

An conventional laser emits a Gaussian beam profile,
$W(\vec{r}) \propto
\linebreak[4]
\exp\left(-2 \vec{r}^2/w^2\right)$, with beam waist~$w$,
which implies a Gaussian filter function
$|W(\vec{q})|^2 \propto \exp\left(-q^2 w^2/4\right)$.
Usually only a small fraction of the molecules is labeled, and then
 $S(\vec{q},t)$ reduces to the incoherent intermediate scattering function
\begin{equation}
S(\vec{q},t) \approx F(\vec{q},t) =
\expect{\exp\big(\text{i} \vec{q} \dotprod \Delta \vec{R}(t)\big)} \, .
\label{eq:inc_isf}
\end{equation}
Considering the displacements $\Delta \vec{R}(t) := \vec{R}(t)-\vec{R}(0)$ after a fixed time lag a random variable, the incoherent scattering function can be interpreted as their characteristic function.
For Gaussian and isotropic displacements, $\expect{\Delta \vec{R}(t)} =0$,
only the second cumulant $\delta r^2(t) := \expect{|\Delta \vec{R}(t)|^2}$ is non-zero.
Thus $F(\vec{q},t) = \exp\left(- q^2  \delta r^2(t)/4\right)$ for
two-\allowhyphens{}dimensional motion.
The corresponding FCS function is calculated to
\begin{equation}
 G_\text{Gauss}(t;w) = \frac{1}{N} \frac{1}{1+ \delta r^2(t)/w^2} \,.
 \label{eq:fcs_gaussian}
\end{equation}
For normal diffusion, it holds $F(\vec{q},t) = \exp\left(-D q^2 t\right)$, and $G(t)$ attains the simple form of \eq{fcs_normal}. For the case of subdiffusion, $\delta r^2(t) \sim t^\alpha$, and Gaussian spatial displacements as in fractional Brownian motion (FBM), one recovers the conventional expression, \eq{fcs_anomalous}.

In many complex systems, however, the (strong) assumption of Gaussian displacements is not valid and may only serve as an approximation.
This assumption can be tested experimentally by resolving the spatial properties of the particle trajectories.
An exact expression for the FCS function is obtained by combining equations~\eqref{eq:fcs_integrals} and \eqref{eq:inc_isf}. Evaluating the integrals over the wavenumber yields
\begin{equation}
 G(t;w) = N^{-1} \expect{\exp\left(-\Delta \vec{R}(t)^2/w^2\right)} \, ,
\label{eq:moment_generating}
\end{equation}
which is a central result of our work.
Let us emphasise that it does not require any assumptions on the dynamics; corrections may
arise from non-dilute labeling of the molecules and from deviations of the Gaussian
beam profile.
In three-dimensional systems, one should further correct for anisotropies in the confocal volume.
This expression enables new insight in the potential of the FCS technique with consequences for the design of future FCS experiments.
The similarity of the representation of $G(t;w)$ in \eq{moment_generating} with that of $F(k,t)$ in \eq{inc_isf} suggests that FCS encodes important spatial information analogous to scattering methods like photon correlation spectroscopy or neutron spin echo.
In the case of anomalous transport discussed below, we will use it as starting point for the derivation of the scaling properties of $G(t;w)$.
Equation~\eqref{eq:moment_generating} shows that the FCS function $N G(t)$ can be neatly interpreted as the return probability for a fluorescent molecule to be again (or still) in the illuminated area.%
\footnote{%
For sufficiently large time lag, the probability to find the fluorophore at a particular point
within the confocal volume becomes independent of the position. Then, the FCS function can be approximated by the probability of being at or returning to the centre of the confocal volume
after the given time multiplied by the size of the confocal volume.}
As a by-product, it provides a simple description for the efficient evaluation of autocorrelated FCS data in computer simulations, circumventing the evaluation of the rapidly fluctuating fluorescent light intensity.

\paragraph*{Van Hove correlation function.}
The dynamics of a single labeled particle is encoded in the probability distribution of the time-dependent displacements,
$P(\vec r,t)=\expect{\delta(\vec r-\vec R(t)}$; due to rotational symmetry, it actually
depends merely on the magnitude $r=|\vec r|$.
This function is also known as van Hove (self-)correlation
function $G(r,t)$ in the field of liquid dynamics~\cite{Hansen:SimpleLiquids}; to avoid confusion with the FCS function, we follow the notation of Ref.~\citenum{benAvraham:DiffusionInFractals}.
Explicit expressions for $P(r, t)$ exist for many models, but for the dynamics
on percolation clusters only conjectures of the asymptotic scaling behaviour are available.
Let us consider a random walker on the incipient infinite percolation cluster, i.e.,
precisely at the percolation threshold.
Then, the dynamics is characterised by two universal exponents:
the fractal dimension $\df$ and the walk dimension $\dw$.
Let further $t_0$ and $\sigma$ denote the typical microscopic time and length scales,
respectively.
The van Hove function is expected to obey the following scaling law for $r\gg\sigma$ and $t\gg t_0$ (Ref. \citenum{benAvraham:DiffusionInFractals}),
\begin{equation}
P_\infty(r,t) = r^{-d} \widehat{P}_\infty (r t^{-1/\dw}) \,.
\label{eq:vanHove_scaling}
\end{equation}
The subscript $\infty$ indicates that the average is taken only for tracers on the infinite cluster.
During a time $t$, the walker explores regions of linear extension of the order of $R \sim t^{1/\dw}$ .
The probability for larger excursions decreases rapidly (presumably like a stretched exponential), hence we assume $\widehat{P}_\infty(x \gg 1) \to 0$ rapidly.
This property specifies the time evolution of the mean-square displacement and of higher moments.
For the FCS measurements, however, we additionally need the limiting behaviour of the scaling function for small arguments.

\paragraph{Return probability.}
Integrating the van Hove function over distances $r\leq w$ with $w$ much larger than any microscopic length yields the probability $\Pi(t;w)$ to return to the starting point of the random walk within a radius~$w$ after a time~$t$.
Provided that $w \ll t^{1/\dw}$, this probability is proportional to the accessible part of the illuminated area, which scales as $w^{\df}$.
In particular, we expect that space- and time-dependence factorise,
\begin{equation}
\Pi(t;w)=\int_{r\leq w}\diff^d\! r\, P_\infty(r,t) \sim w^{\df} \Pi_0(t),
\end{equation}
where $\Pi_0(t)$ denotes the return probability to an infinitesimal vicinity of the origin.
By the scaling law \eq{vanHove_scaling}, we require that
$\widehat{P}_\infty(x\ll 1) \sim x^{\df}$,
which is confirmed by our simulations for the two-\allowhyphens{}dimensional Lorentz model.
As a by-product, one obtains
$\Pi_0(t) \sim t^{-\df/\dw} = t^{-\ds/2}$,
where $\ds = 2 \df/\dw$ is the \emph{spectral dimension}.
Combining both results, $\Pi(t;w) \sim (w t^{-1/\dw})^\df$
for sufficiently long times.

\section*{Models of anomalous transport}

\begin{figure}[t]
\includegraphics[width=3.3in]{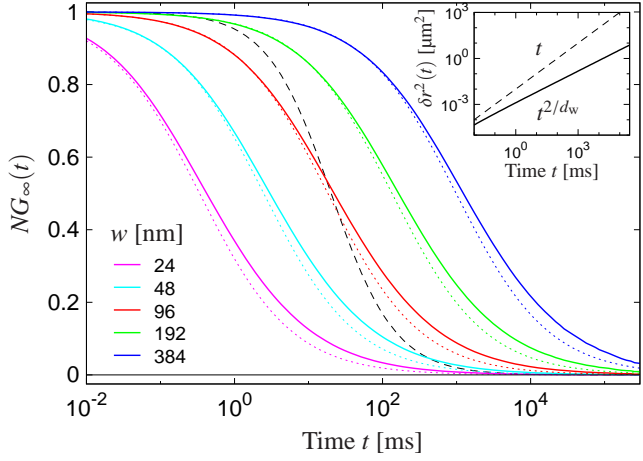}
\caption{Simulated FCS correlation function $G_\infty(t)$ on a logarithmic time axis  for tracers on the infinite cluster at the critical obstacle density;
the beam waist  $w$ of the laser increases from left to right.
Dotted lines correspond to FBM with identical mean-square displacement, \eq{fcs_gaussian};
the broken line displays the FCS function for unobstructed, normal diffusion ($w=450\,\text{nm}$).
Inset: subdiffusive behaviour of the mean-square displacement for the simulated obstructed motion
(solid line), again compared to the case of normal diffusion (broken line).
}
\label{fig:FCS_infty}
\end{figure}

\paragraph*{The Lorentz model.}

Anomalous transport emerges non-trivially in the Lorentz
model~\cite{Lorentz_PRL:2006, Lorentz_LTT:2007, Lorentz_JCP:2008, Lorentz_2D:2010}.
Here, a two-dimensional variant is used which consists of Brownian tracer particles
exploring a disordered environment of randomly placed, overlapping circular obstacles
of radius~$\sigma$, which we choose as $\sigma=3$\,nm.
The void space between the discs undergoes a continuum percolation transition at the critical obstacle density $n_c \sigma^2 \approx 0.35907$ (Ref.~\citenum{Quintanilla:2000}). The infinite cluster displays self-similar behaviour characterised by the fractal dimension $\df=91/48$, known from lattice percolation~\cite{benAvraham:DiffusionInFractals}. The tracer dynamics on this incipient infinite cluster is found to exhibit subdiffusion, $\delta r^2_\infty(t)\sim t^{2/\dw}$, with walk dimension $\dw\approx 2.878$ \mbox{(Ref.~\citenum{Percolation_EPL:2008}),} see inset of \fig{FCS_infty}.

We have generated 1,600 trajectories of Brownian tracers with short-time diffusion coefficient $D_0=2.5$\,\textmu{}m\textsuperscript{2}/s,
moving on the infinite cluster at criticality.
(In practice, we computed trajectories for particles on all clusters
and evaluated the time-averaged mean-square displacement for each particle.
Then, we selected those particles which did not show localisation
based on a criterion for the local exponent of the mean-square discplacement at very long times;
only these particles contributed to the final average over independent trajectories.)
Taking the divergent length scale into account, we have considered large systems of box length $L=10^4\sigma=30$\,\textmu{}m and have run the trajectories up to times of  $t \simeq 10^8 t_0$, where $t_0 = \sigma^2/D_0 = 3.6\,$\textmu{}s is the natural time scale above which the diffusive motion is hindered by obstacles.
The resulting correlation functions are invariant under time shift and do not display aging, in agreement with recent FCS experiments on crowded fluids~\cite{Szymanski:2009}.
For the \emph{in-silico} experiment, we have evaluated the average in \eq{moment_generating} for beam waists between 24\,nm and 384\,nm.

\paragraph*{Fractional Brownian motion.}
Fractional Brownian motion (FBM) is a mathematical generalisation of the usual
Brownian motion yielding a subdiffusive mean-square displacement,
$\delta r^2(t)=2 d D_\alpha t^\alpha$,
with the generalised diffusion constant $D_\alpha$;
the distribution of the displacements $\Delta \vec R(t)$ remains Gaussian.
The description of a microscopic process generating such a dynamics is challenging,
one formulation involving fractional derivatives was given in terms of
a generalised Langevin equation~\cite{Sebastian:1995}.
Nevertheless, its ``propagator'' (van Hove function) can be calculated exactly to
\begin{equation}
 P_\text{FBM}(r,t)=r^{-d}\,
\widehat P_\text{Gauss}\left(r t^{-\alpha/2}/\sqrt{D_\alpha}\right)\,
\end{equation}
where $\widehat P_\text{Gauss}(x)=(2\pi)^{-d/2}x^d\exp\bigl(-x^2/2\bigr)$
and $d$ denotes the dimension of space.
In particular, it satisfies the scaling form in \eq{vanHove_scaling} exactly.
Brownian motion with normal diffusion is obtained in the limit $\alpha\to 1$,
where $D_\alpha$ becomes the diffusion constant.
The FCS function corresponding to FBM is given exactly by \eq{fcs_gaussian}.
For comparison with the Lorentz model, we have fixed $\alpha$ and $D_\alpha$
such that the mean-square displacements of both models coincide.

\section*{Results and Discussion}

In the following, we will describe how FCS experiments with variable beam waist can provide insight into the microscopic dynamics and reveal spatially non-Gaussian, subdiffusive behaviour.
We apply the generalised FCS theory from above to the exactly solvable FBM model and to
the two-dimensional Lorentz model with Brownian tracers.
We have generated FCS correlation functions as described in the previous section.
The obtained curves are shown in \fig{FCS_infty} and exhibit a
significantly stretched  decay compared to normal diffusion.
For the corresponding FBM model with identical mean-square displacement, the same trend, but a different shape of $G(t)$ is found.
For both models, an increase of the beam waist $w$ shifts the relaxation to later times,
while the shape appears to be preserved.

Generalising the diffusion time in \eq{fcs_normal}, we introduce the
half-value time $\tau_{1/2}(w)$ as a function of the beam waist via the implicit
definition $N G(\tau_{1/2})=1/2$.
The FCS data suggest a phenomenological scaling property,
$N G(t;w)=\widetilde G\boldsymbol(t/\tau_{1/2}(w)\boldsymbol)$, i.e., all curves
can be collapsed by appropriate rescaling of time. In the following, we will rigorously
derive the scaling form of the FCS function $G(t;w)$ for the models under consideration.
In particular, a thorough scaling analysis can discriminate whether or not a
proposed theoretical model describes the spatio-temporal tracer dynamics
contained in the FCS data.

\subfiguresbegin
\begin{figure}[t]
\includegraphics[width=3.3in]{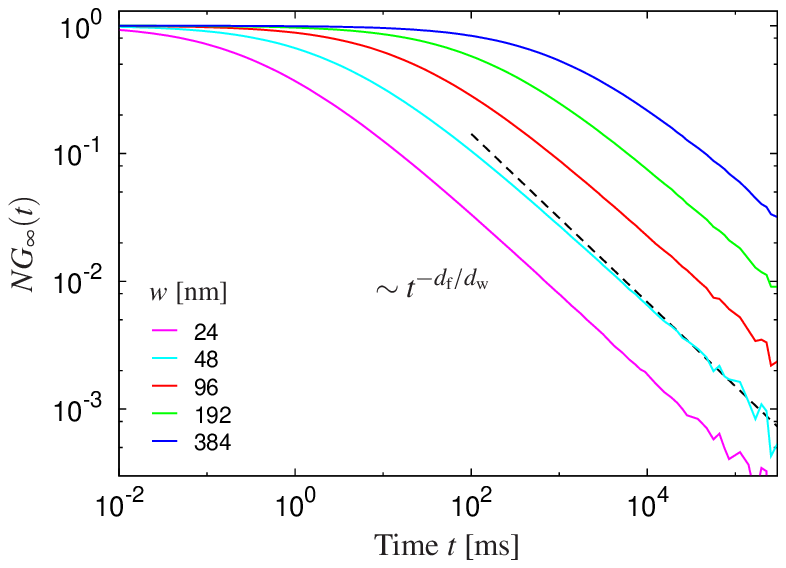}
\caption{Simulated FCS data on double logarithmic scales; the data correspond to \fig{FCS_infty}.
The subdiffusive motion is mainly hidden in the tails of $G_\infty(t;w)$ at long times, $t/t_0 \gg (w/\sigma)^\dw$, which decay as a power law with exponent $\alpha=\df/\dw$, indicated by the broken line. }
\label{fig:FCS_infty_log}
\vspace{\textfloatsep}
\includegraphics[width=3.3in]{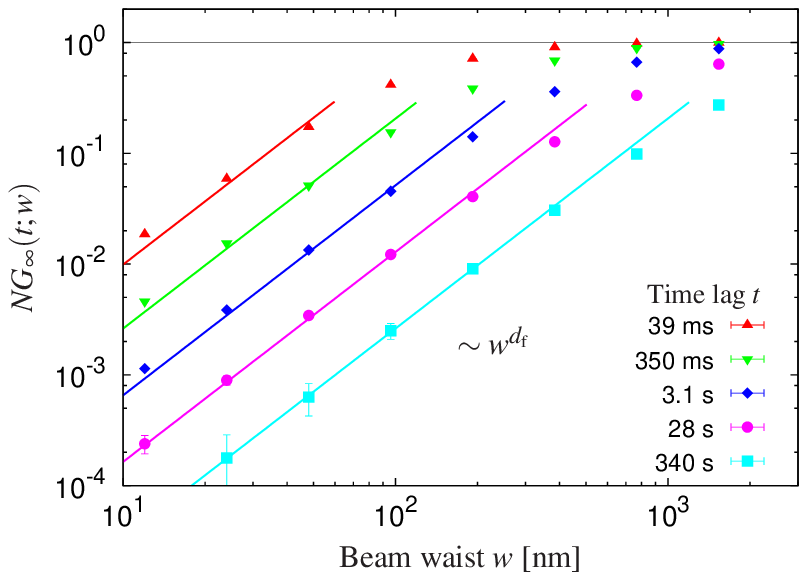}
\caption{Simulated FCS data as a function of the beam waist $w$ for different, but fixed time lags $t$; the data correspond to \fig{FCS_infty}.
The regime $1 \ll w/\sigma \ll (t/t_0)^{1/\dw}$ is characterised by a power law increase with exponent $\df$ (solid lines), revealing the fractal spatial structure.}
\label{fig:FCS_waist-crit}
\end{figure}
\subfiguresend

\subsection*{Scaling of the FCS function.}
The scaling properties of the FCS function are inherited from the van Hove function
by means of the master formula, \eq{moment_generating},
\begin{equation}
N G_\infty(t;w) = \int\! \diff^d r\,  \exp\left(-r^2/w^2\right) P_\infty(r,t)\,.
\end{equation}
In case of the Lorentz model, one finds from \eq{vanHove_scaling} that
\begin{equation}
N G_\infty(t;w) = \widehat{G}_\infty(wt^{-1/\dw})
\label{eq:fcs_scaling}
\end{equation}
for $w\gg \sigma$ and $t\gg t_0$, and similarly for the FBM model.
For both models, these scaling forms imply a power-law divergence
of the FCS half-value time in particular,
\begin{equation}
\tau_{1/2}(w)\sim w^\dw \quad \text{for} \quad w\gg \sigma \,,
\label{eq:half_value_time}
\end{equation}
which is corroborated by the rescaling of $G(t;w)$ below.
For \emph{normal} diffusion on a mesh grid model, a corresponding relation
has been derived~\cite{Destainville:2008}.
In the regime $w\ll t^{1/\dw}$, the FCS experiment essentially probes the return probability $\Pi(t;w)$.
Thus $\widehat{G}_\infty(x\ll 1)\sim x^\df$, and a non-trivial power-law decay of the FCS function is predicted at long times,
\begin{equation}
G_\infty(t;w) \sim t^{-\df/\dw}\,.
\label{eq:fcs_decay}
\end{equation}
A double-logarithmic representation of our simulated FCS data indeed renders the final decay of $G_\infty(t;w)$ straight lines,
see \fig{FCS_infty_log}. Different beam waists yield parallel lines, and the slopes are compatible with the expected value of $\df/\dw=0.66$.

\subfiguresbegin
\begin{figure}
\includegraphics[width=3.3in]{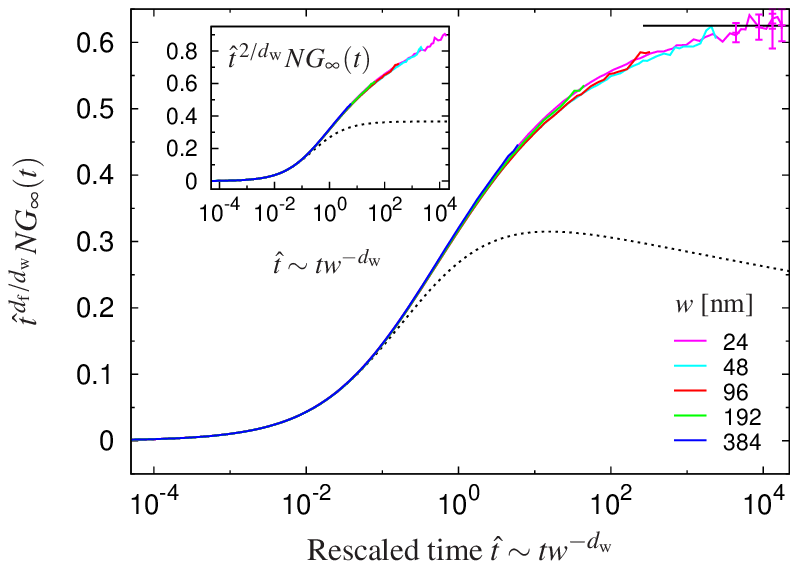}
\caption{FCS functions $G_\infty(t)$ for various beam waists
rescaled according to the scaling hypothesis,
Eqs. \eqref{eq:fcs_scaling} and \eqref{eq:fcs_decay}.
Solid lines correspond to the simulation data of \fig{FCS_infty}, dotted lines to the solution for FBM.
Inset: The assumption of Gaussian transport, \eq{fcs_gaussian}, yields data collapse as well,
but no saturation for large rescaled times.
}
\label{fig:FCS_infty_rescaled}
\vspace{.5\textfloatsep}
\includegraphics[width=3.3in]{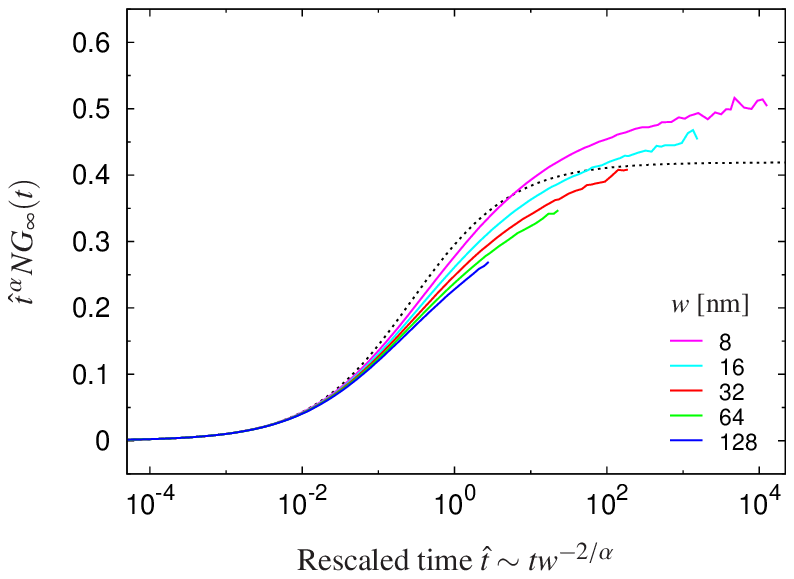}
\caption{The assumption of spatially Gaussian transport, \eq{fcs_gaussian}, does not
lead to scaling of the FCS data for obstructed motion.
The anomalous exponent $\alpha = 0.659$ is obtained from the long-time decay of $G_\infty(t)$
following \eq{fcs_anomalous}. Dotted line: solution for the FBM model rescaled by the same procedure.}
\label{fig:FCS_infty_rescaled_gaussian}
\end{figure}
\subfiguresend

The form of $\widehat G_\infty(\cdot)$ for small arguments further yields the fractal space dimension, $N G_\infty(t;w)\sim w^\df$ for sufficiently large, fixed time lag~$t$.
Thus, the structural properties can be obtained from the FCS data by fixation
of the correlation time and sufficient variation of the beam waist.
and the fractal dimension is directly accessible if the data are plotted on
double-logarithmic scales, see \fig{FCS_waist-crit}.
The asymptotic regime is limited from below by the microscopic details of the system and from above by the crossover to the trivial behaviour, $N G_\infty(t;w\to\infty)=1$.
Within the remaining window, $1 \ll w/\sigma \ll (t/t_0)^{1/\dw}$, our FCS data reveal non-trivial power-law behaviour over 1.5 decades in space for the longest time lags, and the slope of the data corresponds to the fractal dimension $\df=1.9$ of the underlying space.

A simultaneous test of both the temporal decay of $G(t;w)$ and its dependence on the beam waist is provided by appropriate rescaling of the data.
Figure~\ref{fig:FCS_infty_rescaled} shows the FCS function divided by the predicted
long-time decay as a function of the rescaled time $\hat{t} = (t/t_0) (w/\sigma)^{-\dw}$.
The excellent data collapse strongly corroborates the scaling law of \eq{fcs_scaling}
and of the half-value time, \eq{half_value_time}.
Simultaneously, the rectified data converge for $\hat{t}\to \infty$, validating  the algebraic decay, \eq{fcs_decay}.

This analysis is in stark contrast with the form of $G(t)$ obtained for subdiffusive motion and
the assumption of spatially Gaussian transport, e.g., for the FBM model.
It is instructive to discuss the implications for general dimension~$d$.
Then, \eq{fcs_gaussian} generalises to
\begin{equation}
N G^\text{Gauss}_\infty(t;w) = \left[1+ \delta r_\infty^2(t)/w^2 \right]^{-d/2}
\sim w^d t^{-d/\dw}
\label{eq:fcs_gaussian_scaling}
\end{equation}
as $t\to\infty$, using $\delta r^2_\infty(t)\sim t^{2/\dw}$.
The inset of \fig{FCS_infty_rescaled} reveals that rectification with $\hat t^{\,2/\dw}$ does not lead to saturation at long times, reflecting
the fact that the decay of the FCS autocorrelation function for the obstructed motion decays with a different exponent than the inverse mean-square displacement.
Nevertheless, the data still collapse on a single master curve, yet this shows merely that both the Gaussian ansatz and the critical scaling use the same reduced time $\hat{t}$.
Conversely for the FBM result, Gaussian scaling yields convergence at long times, while the assumption of obstructed motion does not.

Let us perform the standard analysis for anomalous transport on our simulated FCS data, as suggested by \eq{fcs_anomalous}.
Fitting the long-time decay fixes the exponent to $\alpha = \df/\dw$, which
indeed yields a saturation in the rectification plot, see \fig{FCS_infty_rescaled_gaussian}. Interpreting this $\alpha$ as characteristic exponent of subdiffusion for the mean-square displacement, $\delta r_\infty^2(t) \sim t^\alpha$, and assuming Gaussian transport, \eq{fcs_gaussian},
suggests the use of $\hat{t} \sim t w^{-2/\alpha}$ as scaling variable.
However, then the data for our model system fan out for large times.
Furthermore, it implies $\tau_{1/2}(w)\sim w^{2/\alpha}=w^{2\dw/\df}$, contradicting
\eq{half_value_time}.
We conclude that the standard approach using only a single scaling exponent
is not consistent for anomalous transport due to obstacles.

We close with the question when the widely used \eq{fcs_anomalous} is a valid
description of the FCS correlation function.
The equation is a specialisation of \eq{fcs_gaussian},
which holds if and only if the distribution of displacements is Gaussian,
i.e., solely determined by its second cumulant, $\delta r^2(t)$; this is a consequence
of the master formula, \eq{moment_generating}.
For spatially Gaussian transport, validity of \eq{fcs_anomalous} is then equivalent
to a power-law increase of the mean-square displacement, $\delta r^2(t) \sim t^\alpha$.
If the functional form of $\delta r^2(t)$ is different, e.g, if the
dynamics exhibits a crossover from anomalous to normal diffusion at some
crossover time scale $t_x$, \eq{fcs_anomalous} applies only to shorter time lags, $t\ll t_x$.
Since such a crossover is generically expected away from a critical point
(see, e.g., the discussion in \mbox{Ref.~\citenum{Avidin:2010}} and the simulation results
in Refs.~\citenum{Lorentz_PRL:2006, Lorentz_JCP:2008, Lorentz_LTT:2007, Lorentz_2D:2010} and   \citenum{Percolation_EPL:2008}),
the analysis of FCS data based on \eq{fcs_gaussian} appears more robust.
Finally as a test of the Gaussian assumption, it would be essential to quantify
the corrections to \eq{fcs_gaussian} by FCS experiments with variable beam waist.

\section*{Conclusions}

We have shown that by systematic variation of the beam waist in FCS experiments, spatio-temporal information on the single-particle dynamics of complex systems can be collected.
We have generalised the FCS theory beyond the assumption of spatially Gaussian transport
and have derived a fundamental expression for the FCS correlation function,
\eq{moment_generating},
which is a general starting point for the interpretation of experiments and which significantly facilitates theoretical and numerical work on FCS.
In particular, it is straightforward to transfer our findings to the study
of complex transport in other fields where FCS is widely employed, e.g., in physical
chemistry and in polymer physics.

The obtained master formula for FCS reveals an analogy between FCS and time-resolved scattering techniques. It can be extended to the case where the concentration of the labeled particles is not dilute any more. Then a  distinct part arises in addition to the self-part similar to the corresponding decomposition of the coherent intermediate scattering function. Likewise, one can easily account for the asphericity of the illuminated volume; yet this does not affect the scaling arguments presented in this work.

For subdiffusive motion due to obstacles, both the fractal nature of the underlying space and the anomalous transport can be revealed by FCS.
We have developed a scaling  theory for $G_\infty(t;w)$,
which excellently describes our simulated data for the full range of investigated beam waists.
These findings have been contrasted to fractional Brownian motion (FBM), an exactly solvable model for subdiffusion with different predictions for the scaling behaviour.
The derived scaling properties should be experimentally accessible with
modern nanoscopic optical methods~\cite{Hell:2007,Eggeling:2009}.
In particular, the spatial information provided by FCS can be used to experimentally distinguish different routes to anomalous transport.

\begin{figure}
\centering
\includegraphics[width=3.3in]{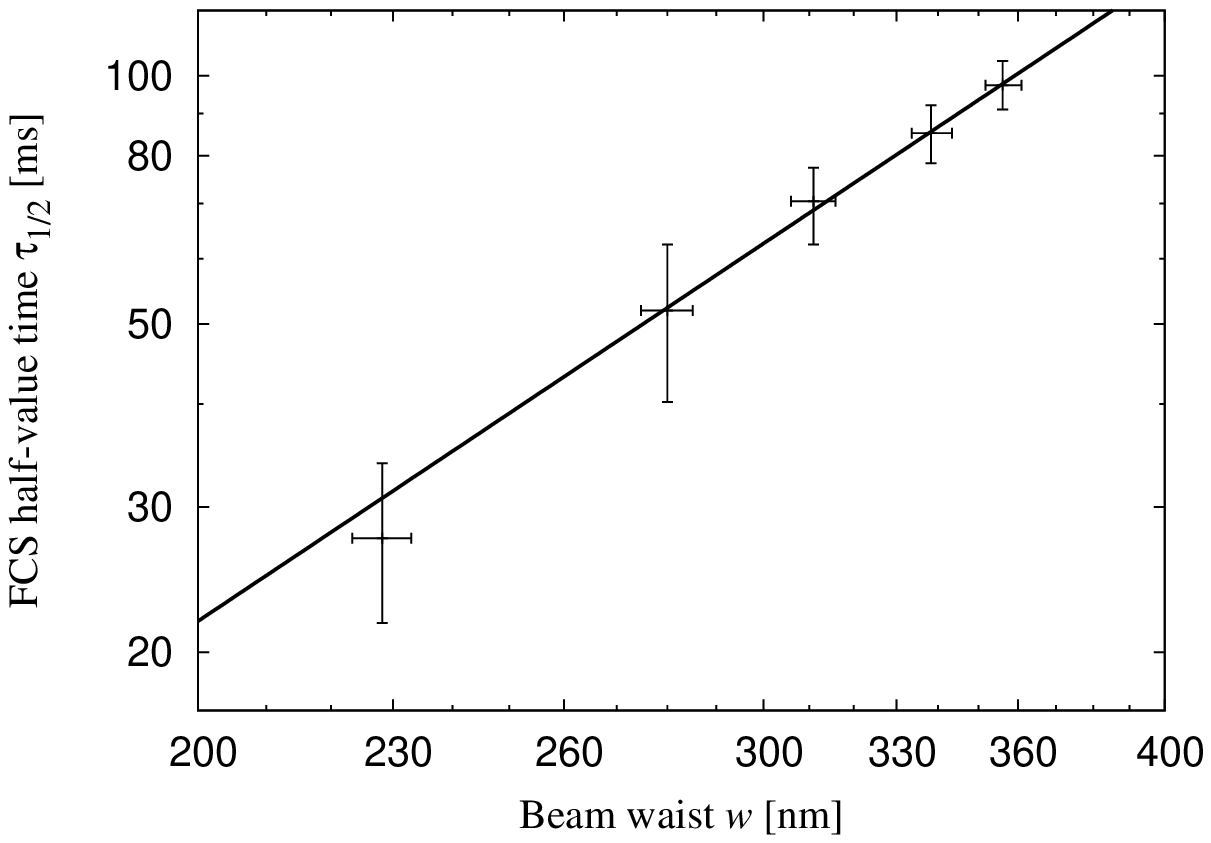}
\caption{Experimental FCS half-value times $\tau_{1/2}(w)$ from FCS measurements
on the transmembrane protein GfR-GFP of COS--7 cells
in a double-logarithmic representation;
the data were taken from Fig.~4D of Ref.~\citenum{Wawrezinieck:2005}.
The straight line indicates a power-law increase of the half-value times as function of beam waist,
$\tau_{1/2}\sim w^{2.6}$.}
\label{fig:wawrezinieck}
\end{figure}

We have demonstrated that a fit of the time-dependence of the FCS function for a single
beam waist does not necessarily determine the walk dimension,
which characterises the subdiffusive increase of the mean-square displacement,
$\delta r^2(t)\sim t^{2/\dw}$.
A more robust procedure would be based on measurements of the half-value time $\tau_{1/2}(w)$
of the normalised FCS correlations for a wide range of beam waists. Then, the exponent
of subdiffusion may be obtained from the $w$-dependence
$\tau_{1/2}(w)\sim w^\dw$,
which is expected to hold for various models of anomalous transport, see \eq{half_value_time}.
The knowledge of $\dw$ would be the starting point for a scaling test of the
full curves $G(t;w)$ similarly to \fig{FCS_infty_rescaled}, which then would be completed
by a characterisation of the decay at long times.
The analysis by Wawrezinieck \emph{et al.} \cite{Wawrezinieck:2005} is somewhat different from the
one suggested here. Their ``apparent diffusion time'' $\tau_\text{d}^\text{app}$ corresponds to $\tau_{1/2}$ in our notation. Based on experimental data on the plasma membrane of COS--7 cells and on simulations, they find the phenomenological relation $\tau_{1/2}\sim w^2 + \mathit{const}$
and discuss implications of the offset as the data are extrapolated towards
small beam waists, $w\to 0$.
In \fig{wawrezinieck}, we have replotted their $\tau_{1/2}$-data for the
transmembrane protein GfR-GFP as function of the beam waist on double-logarithmic
scales. The data points nicely follow a straight line in agreement with our
power-law prediction, \eq{half_value_time}.
Considering the error bars and the limited $w$-range, we obtain an estimate of the walk
dimension $\dw$ between 2.4 and 3.4; the most likely value is $\dw\approx2.6$,
implying an exponent of subdiffusion of $2/\dw\approx 0.77$.
We find this observation rather encouraging
with respect to the applicability of our approach and to the usefulness of FCS
with variable beam waist for addressing the leading questions on anomalous transport.

\paragraph{Acknowledgment.}
We thank J.~O. R\"adler for insightful discussions on the experimental techniques.
Financial support  from the Deutsche Forschungsgemeinschaft via contract No. FR \mbox{850/6--1}
and by the  German Excellence Initiative via the program ``Nanosystems Initiative
Munich'' (NIM) is gratefully acknowledged.


\end{document}